\documentclass[12pt,runningheads]{llncs}

\usepackage[utf8]{inputenc}    
\usepackage{amsmath}
\usepackage{amssymb}
\usepackage{graphicx}
\usepackage{epstopdf}
\usepackage{inputenc}
\usepackage{geometry}
\usepackage{mdframed}

\usepackage{float}

\usepackage{authblk}
\usepackage{url}
\usepackage{parskip}
\usepackage{authblk}
\usepackage{color}

    \usepackage[colorinlistoftodos,prependcaption,textsize=tiny]{todonotes}
    \usepackage[bookmarks, colorlinks, breaklinks]{hyperref}
\makeatletter
\def\blfootnote{\xdef\@thefnmark{}\@footnotetext}
\makeatother

\providecommand{\tightlist}{%
  \setlength{\itemsep}{0pt}\setlength{\parskip}{0pt}}

\hypersetup
{
  pdfauthor={Philip B. Stark},
  pdfsubject={election audits, ballot-marking devices, election integrity},
  pdftitle={Sets of Half-Average Nulls Generate Risk-Limiting Audits: SHANGRLA},
  colorlinks,breaklinks,
  filecolor=black,
  urlcolor=[rgb]{0.117,0.682,0.858},
  linkcolor=[rgb]{0.117,0.682,0.858},
  linkcolor=[rgb]{0.117,0.682,0.858},
  citecolor=[rgb]{0.117,0.682,0.858}
}

\begin{document}

\title {Sets of Half-Average Nulls Generate Risk-Limiting Audits: SHANGRLA}

\author{
Philip B. Stark
}

\institute{
University of California, Berkeley
}

\maketitle

\abstract{Risk-limiting audits (RLAs) for many social choice functions can be
reduced to testing sets of null hypotheses of the form ``the average of
this list is not greater than 1/2'' for a collection of finite lists of
nonnegative numbers. Such social choice functions include majority,
super-majority, plurality, multi-winner plurality, Instant Runoff Voting
(IRV), Borda count, approval voting, and STAR-Voting, among others. The
audit stops without a full hand count iff all the null hypotheses are
rejected. The nulls can be tested in many ways. Ballot polling is
particularly simple; two new ballot-polling risk-measuring functions for
sampling without replacement are given. Ballot-level comparison audits
transform each null into an equivalent assertion that the mean of
re-scaled tabulation errors is not greater than 1/2. In turn, that null
can then be tested using the same statistical methods used for ballot
polling---applied to different finite lists of nonnegative numbers: the
SHANGRLA approach reduces auditing different social choice functions and
different audit methods to the same simple statistical problem.
Moreover, SHANGRLA comparison audits are more efficient than previous
comparison audits for two reasons: (i) for most social choice functions,
the conditions tested are both necessary and sufficient for the reported
outcome to be correct, while previous methods tested conditions that
were sufficient but not necessary, and (ii) the tests avoid a
conservative approximation. The SHANGRLA abstraction simplifies
stratified audits, including audits that combine ballot polling with
ballot-level comparisons, producing sharper audits than the ``SUITE''
approach. SHANGRLA works with the ``phantoms to evil zombies'' strategy
to treat missing ballot cards and missing or redacted cast vote records.
That also facilitates sampling from ``ballot-style manifests,'' which
can dramatically improve efficiency when the audited contests do not
appear on every ballot card. Open-source software implementing SHANGRLA
ballot-level comparison audits is available. SHANGRLA was tested in a
process pilot audit of an instant-runoff contest in San Francisco, CA, in
November, 2019.}
\\[3em] \textbf{Keywords}: sequential tests, martingales, Kolmogorov's inequality

\textbf{Acknowledgments:} I am grateful to Andrew Conway, Steven N.
Evans, Kellie Ottoboni, Ronald L. Rivest, Vanessa Teague, Poorvi Vora,
and Damjan Vukcevic for helpful conversations and comments on earlier
drafts, and to Filip Zagorski for presenting the paper at the 5th
Workshop on Advances in Secure Electronic Voting. The SHANGRLA software
was a collaborative effort that included Michelle Blom, Andrew Conway,
Dan King, Laurent Sandrolini, Peter Stuckey, and Vanessa Teague.

\hypertarget{sec:intro}{%
\section{Introduction}\label{sec:intro}}

A \emph{risk-limiting audit} (RLA) of a reported election contest
outcome is any procedure that guarantees a minimum probability of
correcting the reported outcome if the reported winner(s) did not really
win, but never alters a correct reported outcome. The largest
probability that the procedure fail to correct the reported outcome if
the reported outcome is wrong is the \emph{risk limit}.

RLAs were introduced by \cite{stark08a} and named by \cite{stark09b}.
RLA methods have been developed for a variety of social choice
functions, to accommodate election equipment with different
capabilities, and to comport with the logistics of ballot handling,
organization, and storage in different jurisdictions.

RLAs are considered the gold standard for post-election tabulation
audits, recommended by the National Academies of Science, Engineering,
and Medicine \cite{NASEM18}, the Presidential Commission on Election Administration
\cite{PCEA14}, the American
Statistical Association \cite{asa10}, the
League of Women Voters, Verified Voting Foundation, the Brennan Center
for Justice, and other organizations concerned with election integrity.

\paragraph{Experience with RLAs.}
RLAs have been piloted dozens of times in 11 U.S. states and in Denmark.
They are required by statute in Colorado, Nevada, Rhode Island, and
Virginia, and authorized by statute in California and Washington.

\paragraph{Resources for RLAs.}
There is free and open-source software to help with the random selection
of ballot cards for RLAs and to perform the risk calculations to determine when
and if the audit can stop.\footnote{See, e.g.,
  \url{https://www.stat.berkeley.edu/users/stark/Vote/auditTools.htm},
  \url{https://www.stat.berkeley.edu/users/stark/Vote/ballotPollTools.htm},
  \url{https://github.com/pbstark/auditTools},
  \url{https://github.com/pbstark/CORLA18/blob/master/code/suite\_toolkit.ipynb},
  and \url{https://github.com/votingworks/arlo} (all last visited 10 November
  2019); an implementation of SHANGRLA ballot-level comparison audits is
  available at \url{https://github.com/pbstark/SHANGRLA}.}

\paragraph{Prerequisites for RLAs.}
A risk-limiting audit of a trustworthy paper record of voter intent can
catch and correct wrong election outcomes. (Here, ``trustworthy'' means
that a full hand count of the votes in the record would show the true
winners.) But RLAs themselves check only the tabulation, not the
trustworthiness of the paper records. If the paper trail is not
trustworthy, it is not clear what a risk-limiting audit procedure
accomplishes: while it can offer assurances that \emph{tabulation error}
did not alter the reported outcome, it cannot determine whether the
reported outcome is right or wrong, nor can it promise any probability
of correcting wrong outcomes. 
It therefore cannot limit the risk of certifying an
outcome that is incorrect---the ``risk'' that a risk-limiting audit is
supposed to limit.

Because all electronic systems are vulnerable to bugs, misconfiguration,
and hacking, the paper trail can be trustworthy only if elections are
conducted using \emph{voter-verified paper ballots kept demonstrably
secure} throughout the canvass and the audit. In particular,
\emph{compliance audits} 
\cite{benalohEtal11,starkWagner12,lindemanStark12,starkEBE18} 
are needed to assure that the paper
trail remains inviolate from the time of casting through the completion
of the audit.

However, that is not enough: the means of marking the paper matters.
Recent experiments have shown that voters rarely check BMD printout and
rarely notice errors in BMD printout, even when instructed to
check---verbally, through written instructions, and through signage
\cite{bernhardEtal20}. Most BMD output is evidently not voter-verified.
Moreover, while voters who notice errors in BMD printout may request a
fresh chance to mark a ballot, there is no way to tell whether the error
was the voter's fault or the BMD's fault 
\cite{kaczmarekEtal13,bernhardEtal17,appelEtal19}. As a result,
malfunctioning BMDs may go undetected by election officials. Applying
RLA procedures to BMD printout cannot limit the risk that incorrect
reported outcomes will go uncorrected, unless one simply defines
``correct outcome'' to be whatever an accurate tally of the paper would
show, whether or not that reflects what voters indicated to the
equipment.

\paragraph{What's new here.}
SHANGRLA uses a new abstract framing of RLAs that involves constructing
a set of \emph{assertions} for each contest. 
If the assertions are true, the contest outcomes are correct.
The assertions are
predicates on the set of ballot cards, that is, they are either true or
false, depending on what votes the whole set of trusted paper ballot
cards shows.

Each assertion is characterized by an \emph{assorter}, a function that
assigns a nonnegative number to each ballot card,\footnote{A ballot
  consists of one or more ballot cards. Below, ``ballot,'' ``card,'' and
  ``ballot card'' are used interchangeably, even though in most U.S.
  jurisdictions, a ballot consists of more than one ballot card.} again
depending on the votes reflected on the ballot. 
The assertions that
characterize the audit are of the form ``the average value of the
assorter for all the cast ballots is greater than \(1/2\).'' In turn,
each assertion is checked by testing the \emph{complementary null
hypothesis} that the average is less than or equal to \(1/2\). To reject
the entire set of complementary null hypotheses is to confirm the
outcomes of all the contests under audit. Hence, the name of this
method: Sets of Half-Average Nulls Generate RLAs (SHANGRLA).

By reducing auditing to repeated instances of a single, simple
statistical problem---testing whether the mean of a list of nonnegative
numbers is less than 1/2---SHANGRLA puts ballot-polling audits,
comparison audits, batch-level comparison audits, and stratified and
``hybrid'' audits on the same footing, and puts auditing a broad
range of social choice functions on the same footing.
Moreover, it makes it easy to incorporate
statistical advances into RLAs: only one function needs to be updated.

Open-source software implementing SHANGRLA audits is
available.\footnote{\url{https://www.github.com/pbstark/SHANGRLA}, last
  visited 22 November 2019.} The software also implements the ``phantoms
to evil zombies'' approach of \cite{banuelosStark12} for dealing with
missing cast-vote records and missing ballot cards.
That also makes it possible to sample from ``ballot-style manifests'' 
\cite{benalohEtal11,lindemanEtal18}, which facilitates efficient
audits of contests that do not appear on every ballot card cast in the
election. Despite the fact that they were developed more than 7~years
ago, neither ``phantoms-to-zombies'' nor sampling from ballot-style
manifests has been implemented before. 
The SHANGRLA software was tested in
practice in a process pilot audit in San Francisco, CA, in November 2019.

\hypertarget{sec:assorter}{%
\section{Assorted Simplifications}\label{sec:assorter}}

An \emph{assorter} \(A\) assigns a nonnegative value to each ballot card,
depending on the marks the voter made on that ballot card.\footnote{%
 The value might also depend on what the voting system reported
 for that ballot card and others. See section~\ref{sec:ballot-comparison}.
 }

For instance, suppose that Alice and Bob are running against each other
in a two-candidate first-past-the-post contest. 
The following function
is an assorter: assign the value ``1'' to a ballot card if it has a mark for
Alice but not for Bob; assign the value ``0'' if the card has a mark for
Bob but not for Alice; assign the value 1/2, otherwise (e.g., if the
card has an overvote or an undervote in this contest or does not contain
the contest). 
Then Alice beat Bob iff the average value of the assorter
for the full set of cast ballot cards is greater than \(1/2\):
then Alice got more than 50\% of the valid votes.

To express this more mathematically, let \(b_i\) denote the \(i\)th
ballot card, and suppose there are \(N\) ballot cards in all. Let
\(1_\mathrm{Alice}(b_i) = 1\) if ballot \(i\) has a mark for Alice, and
\(0\) if not; define \(1_\mathrm{Bob}(b_i)\) analogously. The assorter
could be written \[
  A_{\mathrm{Alice, Bob}} (b_i)  \equiv \frac{1_\mathrm{Alice}(b_i) - 1_\mathrm{Bob}(b_i) + 1}{2}.
\] 
If \(b_i\) shows a mark for Alice but not for Bob,
\(A_{\mathrm{Alice, Bob}}(b_i) = 1\). 
If it shows a mark for Bob but not
for Alice, \(A_{\mathrm{Alice, Bob}}(b_i) = 0\). If it shows marks for
both Alice and Bob (an overvote), for neither Alice nor Bob (an
undervote), or if the ballot card does not contain the Alice v. Bob
contest at all, \(A_{\mathrm{Alice, Bob}}(b_i) = 1/2\). The average
value of \(A\) over all ballot cards is \[
   \bar{A}_{\mathrm{Alice, Bob}}^b \equiv \frac{1}{N} \sum_{i=1}^N A_{\mathrm{Alice, Bob}}(b_i).
\]

If Alice is the reported winner, the contest can be audited at risk
limit \(\alpha\) by testing the \emph{complementary null hypothesis}
that \(\bar{A}_{\mathrm{Alice, Bob}}^b \le 1/2\) at significance level
\(\alpha\). To reject the complementary null hypothesis is to conclude
that Alice really won. If the complementary null hypothesis is true, Bob
won or the contest was a tie: the assertion is false.

An assorter offers more flexibility than just counting votes.
For instance, instead of either giving Alice one vote, half a vote, or
no vote, an assorter could interpret a ballot card as giving Alice an
arbitrary nonnegative value as a vote, depending on the marks on the
ballot. This flexibility lets assorters solve the problem of auditing
more complicated social choice functions, as we shall see.

\hypertarget{sec:plurality}{%
\subsection{Plurality elections}\label{sec:plurality}}

In a plurality contest with \(K \ge 1\) winners and \(C>K\) candidates
in all, a collection of candidates \(\{w_k\}_{k=1}^K\) are the true
winners and the remaining \(C-K\) candidates \(\{\ell_j\}_{j=1}^{C-K}\)
are the true losers iff the assertions

\[
  \bar{A}_{w_k, \ell_j}^b > 1/2,  \;\; \mbox{ for all } 1 \le k \le K, \;\; 1 \le j \le C-K
\] all hold, where \(A_{w_k, \ell_j}\) is defined as above, with \(w_k\)
taking the role of Alice and \(\ell_j\) taking the role of Bob. 
(This is
essentially the approach taken in 
\cite{stark08a,stark09a,stark10d,lindemanStark12}, reformulated and in different notation.) 
The
contest can be audited to risk limit \(\alpha\) by testing the
\(K(C-K)\) hypotheses \(\bar{A}_{w_k, \ell_j}^b \le 1/2\) individually
at significance level \(\alpha\). The audit stops only if all \(K(C-K)\)
complementary hypotheses are rejected; otherwise it requires a full hand
count.

\hypertarget{sec:approval}{%
\subsection{Approval Voting}\label{sec:approval}}

Even though the voting rules are different, the same assorter functions
can be used to audit approval voting and plurality voting. Candidates
\(\{w_k\}_{k=1}^K\) are the winners and the remaining \(C-K\) candidates
\(\{\ell_j\}_{j=1}^{C-K}\) are the losers of a $K$-winner approval voting contest 
iff all the assertions \[
  \bar{A}_{w_k, \ell_j}^b > 1/2,  \;\; \mbox{ for all } 1 \le k \le K, \;\; 1 \le j \le C-K
\] hold.

\hypertarget{sec:super-majority}{%
\subsection{Super-majority}\label{sec:super-majority}}

Suppose that a candidate must get at least a fraction \(f \in (0, 1)\)
of the valid votes to win.\footnote{Values \(f \le 1/2\) are not
  technically ``super-majorities,'' but the generality is useful. For
  instance, the rules of some primaries in the U.S. eliminate candidates
  who receive less than 15\% of the vote.
  An RLA using $f = 0.15$ might be used to check whether the correct candidates were eliminated.
} 
\cite{stark08a} shows how to
audit this social choice function, but it can also be expressed in terms
of the assertion that the average of an assorter applied to the cast
ballot cards is greater than \(1/2\).

Alice really won a super-majority contest with required winning fraction
\(f\) iff \[
\mbox{(valid votes for Alice)} > f \times \left ( \mbox{(valid votes for Alice)} + 
  \mbox{(valid votes for everyone else)} \right ).
\]

Define an assorter as follows: \begin{equation} 
  A(b_i) \equiv \left \{ \begin{array}{ll} 
        \frac{1}{2f}, & \mbox{$b_i$ has a mark for Alice and no one else} \\
        0, & \mbox{$b_i$ has a mark for exactly one candidate and not Alice} \\
        1/2, & \mbox{otherwise}.
        \end{array}
        \right .
\label{eq:super-majority-assorter}\end{equation} This assigns a
nonnegative number to every ballot. Suppose that a fraction \(p > f\)
of the valid votes are for Alice, and that a fraction \(q\) of the
ballots have valid votes. Then

\[
 \bar{A}^b \equiv pq/(2f) + (1-q)/2 \ge q/2 + (1-q)/2 = 1/2.
\]

Again, using assorters reduces auditing to the question of whether the
average of a list of nonnegative numbers is greater than 1/2. The
correctness of the outcome is implied by a single assertion, unlike
plurality elections, which require (number of winners)\(\times\)(number
of losers) assertions.\footnote{%
  To check whether $K$ candidates all got at least a fraction $f \in (0, 1)$
  of the valid votes (with $Kf < 1$) requires testing at most $K$ assertions.
}

\hypertarget{sec:dhondt}{%
\subsection{D'Hondt and other proportional representation
schemes}\label{sec:dhondt}}

\cite{starkTeague14} show how to reduce the problem of auditing
D'Hondt and other proportional representation social choice functions to
the problem of auditing a collection of two-candidate plurality contests.
We have seen above that each such two-candidate contest can be expressed
as the assertion that the average of an assorter applied to the ballots
is greater than \(1/2\), so auditing proportional representation
contests can be reduced to auditing a collection of assertions that the
averages of a set of assorters over the cast ballots is greater than
\(1/2\).

\hypertarget{sec:borda}{%
\subsection{Borda count, STAR-Voting, and other weighted additive voting
schemes}\label{sec:borda}}

Borda count is one of several voting systems that assign points to
candidates for each ballot, depending on what the ballot shows; the
winner is the candidate who receives the most points in total across all
cast ballots. This involves only a slight generalization of plurality
contests to account for the fact that a ballot can give a candidate more
than one ``point,'' while for plurality a candidate either gets a vote
or not. As before, the reported result is correct if the reported winner
actually received more points than each reported loser, which we can
test by constructing an assorter for each (winner, loser) pair.

Let \(s_{\mathrm{Alice}}(b_i)\) denote a nonnegative ``score'' for
Alice on ballot \(i\), and let \(s_{\mathrm{Bob}}(b_i)\) be the score
for Bob. These need not be integers. Let \(s^+\) be an upper bound on
the score any candidate can get on a ballot. Alice beat Bob iff

\[
  \sum_{i=1}^N s_{\mathrm{Alice}}(b_i) > \sum_{i=1}^N s_{\mathrm{Bob}}(b_i),
\] i.e., iff \[
    \bar{s}_{\mathrm{Alice}}^b > \bar{s}_{\mathrm{Bob}}^b.
\] Make the affine transformation \[
   A(b_i) \equiv \frac{s_{\mathrm{Alice}}(b_i) - s_{\mathrm{Bob}}(b_i) + s^+}{2s^+}.
\] Then \(A(b_i) \ge 0\) and
\(\bar{s}_{\mathrm{Alice}}^b > \bar{s}_{\mathrm{Bob}}^b\) iff
\(\bar{A}^b > 1/2\).

\hypertarget{sec:IRV}{%
\subsection{Ranked-Choice and Instant-Runoff Voting
(RCV/IRV)}\label{sec:IRV}}

\cite{blomEtal19} show how to reduce the correctness of a
reported IRV winner to the correctness of the reported winners of a set
of two-candidate plurality contests. The ``candidates'' in those contests are not
necessarily the candidates in the original contest; they are just two
mutually exclusive (but not exhaustive) conditions that a ballot might
satisfy.

Two types of assertions can be combined to give
sufficient conditions for the reported winner of an IRV contest to have
really won:

\begin{enumerate}
\def\labelenumi{\arabic{enumi}.}
\tightlist
\item
  Candidate \(i\) has more first-place ranks than candidate \(j\) has
  total mentions.
\item
  After a set of candidates have been eliminated from consideration,
  candidate \(i\) is the top ranked candidate on more ballot cards than
  candidate \(j\) is.
\end{enumerate}

Both of these can be written as \(\bar{A}^b > 1/2\) by labeling the
corresponding vote patterns ``Alice'' or ``Bob'' or ``neither.''

For instance, consider the first type of assertion. If \(b_i\) has
candidate \(i\) ranked 1, the ballot is considered a vote for Alice. If
\(b_i\) ranks candidate \(j\) at all, the ballot is considered a vote
for Bob. Otherwise, the ballot is not a vote for either of them. If
Alice beat Bob, candidate \(j\) cannot have beaten candidate \(i\) in
the IRV contest.

In contrast to plurality, supermajority, approval, Borda, and d'Hondt,
the assertions derived by \cite{blomEtal19} are
sufficient for the reported winner to have won, but not necessary.
Hence, it might be possible to sharpen such audits.

\hypertarget{sec:auditing}{%
\section{Auditing assertions}\label{sec:auditing}}

We audit the assertion \(\bar{A}^b > 1/2\) by testing the
\emph{complementary null hypothesis} \(\bar{A}^b \le 1/2\)
statistically. We audit until either all complementary null hypotheses
about a contest are rejected at significance level \(\alpha\) or until
all ballots have been tabulated by hand. This yields a RLA of the
contest in question at risk limit \(\alpha\).

\hypertarget{sec:ballot-polling}{%
\subsection{Ballot-polling audits}\label{sec:ballot-polling}}

Ballot-polling audits select individual ballot cards at random, either
with or without replacement. The BRAVO method of \cite{lindemanEtal12} 
uses Wald's Sequential Probability Ratio Test (SPRT) for
sampling with replacement.

For each (reported winner, reported loser) pair, BRAVO tests the
conditional probability that a ballot contains a vote for the reported
winner given that it contains a vote for the reported winner or the
reported loser. Using assorters allows us to eliminate the conditioning
and opens the door to a broader collection of statistical tests,
including tests based on sampling \emph{without} replacement, which can
improve the efficiency of the audit. 
Two such methods are presented
below. 
In contrast to the SPRT (with or without replacement), 
these methods only require knowing the
reported winners, not the reported vote shares.

First, we shall derive the Kaplan-Kolmogorov method for sampling without
replacement, based on ideas in Harold Kaplan's (now defunct)
website.\footnote{\url{http://web.archive.org/web/20131209044835/http://printmacroj.com/martMean.htm}}
The method is based on the observation that a suitably constructed
sequence is a nonnegative martingale, to which Kolmogorov's inequality
for optionally stopped closed martingales can be applied.

We sample without replacement from a finite population of \(N\)
nonnegative items, \(\{x_1, \ldots, x_N\}\), with \(x_j \ge 0\),
\(\forall j\). The population mean is
\(\mu \equiv \frac{1}{N} \sum_{j=1}^N x_j \ge 0\) and the population total is
\(N\mu \ge 0\). The value of the \(j\)th item drawn is \(X_j\). On the
hypothesis that \(\mu = t\), \(\mathbb{E}X_1 = t\), so
\(\mathbb{E}(X_1/t) = 1\). Conditional on \(X_1, \ldots, X_n\), the
total of the remaining \(N-n\) items is \(N\mu - \sum_{j=1}^n X_j\), so
the mean of the remaining items is \[
    \frac{Nt-\sum_{j=1}^n X_j}{N-n} = \frac{t - \frac{1}{N} \sum_{j=1}^n X_j}{1-n/N}.
\] Thus, the expected value of \(X_{n+1}\) given \(X_1, \ldots, X_n\) is
\(\frac{t - \frac{1}{N} \sum_{j=1}^n X_j}{1-n/N}\). Define \[
Y_1(t) \equiv
\begin{cases}
X_1/t,& Nt > 0, \\
1,&   Nt = 0, \\
\end{cases}
\] and for \(1 \le n \le N-1\), \[
Y_{n+1}(t) 
\equiv
\begin{cases}
X_{n+1} \cdot
\frac
{1 - \frac{n}{N}}
{t - \frac{1}{N} \sum_{j=1}^n X_j},& \sum_{j=1}^n X_j < Nt, \\
1,& \sum_{j=1}^n X_j \ge Nt. \\
\end{cases}
\] Then \(\mathbb{E}(Y_{n+1}(t) | Y_1, \ldots Y_n) = 1\). Let
\(Z_{n}(t) \equiv \prod_{j=1}^n Y_j(t)\). Note that \(Y_k(t)\) can be
recovered from \(\{Z_j(t), j \le k\}\), since
\(Y_k(t) = Z_k(t)/Z_{k-1}(t)\). Now
\(\mathbb{E}|Z_k| \le \max_j x_j < \infty\) and \[
   \mathbb{E}\left ( Z_{n+1}(t) | Z_1(t), \ldots Z_n(t) \right ) = 
   \mathbb{E} \left (Y_{n+1}(t)Z_n(t) | Z_1(t), \ldots Z_n(t) \right ) = Z_n(t).
\] Thus \[
    \left ( Z_1(t), Z_2(t), \;\ldots , Z_N(t) \right )
\] is a nonnegative closed martingale. By Kolmogorov's inequality, an
application of Markov's inequality to martingales \cite[p242]{Fel71}
for any \(p > 0\) and any \(J \in \{1, \ldots, N \}\), \[
     \Pr \left ( \max_{1 \le j \le J} Z_j(t) > 1/p \right ) \le p \; \mathbb{E}|Z_J|.
\] Since \((Z_j)\) is a nonnegative martingale,
\(\mathbb{E}|Z_J| = \mathbb{E}Z_J = \mathbb{E}Z_1 = 1\). Thus a
\(P\)-value for the hypothesis \(\mu = t\) based on data
\(X_1, \ldots X_J\) is
\(\left (\max_{1 \le j \le J} Z_j(t) \right )^{-1} \wedge 1\).

However, if \(X_j = 0\) for some \(j\), then \(Z_k = 0\) for all
\(k \ge j\). To avoid that problem, we can shift everything to the
right: pick \(\gamma > 0\), find a lower confidence bound for
\(\delta = \mu+\gamma >  0\) from data \(\{X_j+\gamma\}\), then
subtract \(\gamma\) from the lower confidence bound to get a lower
confidence bound for \(\mu\). There are tradeoffs involved in picking
\(\gamma\): if many \(X_j\) turn out to be small, especially for small
\(j\), it helps to have \(\gamma\) large, and vice versa.

Unpacking the math yields the \(P\)-value \[
p_{\mathrm{KK}} \equiv 1 \wedge \left ( \max_{1 \le j \le J} 
\prod_{k=1}^j (X_{k}+\gamma) \frac{1-(k-1)/N}{t - \frac{1}{N} \sum_{\ell=1}^{k-1} (X_\ell+\gamma)} \right )^{-1}
\] for the hypothesis that \(\mu \le t - \gamma\). This is implemented
in the SHANGRLA software.

A related test that uses sampling without replacement, also introduced
without proof on Kaplan's website, can be derived as follows. Let
\(S_j \equiv \sum_{k=1}^j X_k\), \(\tilde{S}_j \equiv S_j/N\), and
\(\tilde{j} \equiv 1 - (j-1)/N\). Define \[
   Y_n \equiv \int_0^1 \prod_{j=1}^n 
   \left ( \gamma \left [ X_j \frac{\tilde{j}}{t - \tilde{S}_{j-1}} -1 \right ] + 1 \right ) d\gamma.
\] This is a polynomial in \(\gamma\) of degree at most \(n\), with
constant term \(1\). Each \(X_j\) appears linearly. Under the null
hypothesis that the population total is \(Nt\), \(\mathbb{E} X_1 = t\),
and \[
  \mathbb{E} \left ( X_j \mid X_1, \ldots, X_{j-1} \right ) = 
  \frac{Nt - S_{j-1}}{N-j+1} = \frac{t - \tilde{S}_{j-1}}{\tilde{j}}.
\] Now \[
   Y_1 = \int_0^1 \left ( \gamma [ X_1/t - 1] + 1 \right ) d\gamma = 
   \left [ (\gamma^2/2) [X_1/t - 1] + \gamma \right ]_{\gamma=0}^1 = 
   [X_1/t - 1]/2 + 1 = \frac{X_1}{2t} + 1/2.
\] Thus, under the null, \[
   \mathbb{E}Y_1 = \frac{\mathbb{E}X_1}{2t} + 1/2 = 1.
\] Also, \[
   \mathbb{E}(Y_n | X_1, \ldots, X_{n-1}) =
   \mathbb{E} \left . \left [ \int_0^1 \prod_{j=1}^n \left (\gamma \left [ X_j \frac{\tilde{j}}{t - \tilde{S}_{j-1}} -1 \right ] + 1 \right ) d\gamma \right | X_1, \ldots, X_{n-1} \right ]
\] \[
  = \int_0^1  \left (\gamma \left [ \mathbb{E}(X_n | X_1, \ldots, X_{n-1}) \frac{\tilde{n}}{t - \tilde{S}_{n-1}} -1 \right ] + 1 \right ) \prod_{j=1}^{n-1} \left ( \gamma \left [ X_j \frac{\tilde{j}}{t - \tilde{S}_{j-1}} -1 \right ] + 1 \right ) d\gamma 
\] \[
  = \int_0^1  \left (\gamma \left [ \frac{t - \tilde{S}_{n-1}}{\tilde{n}} \frac{\tilde{n}}{t - \tilde{S}_{n-1}} -1 \right ] + 1 \right ) \prod_{j=1}^{n-1} \left ( \gamma \left [ X_j \frac{\tilde{j}}{t - \tilde{S}_{j-1}} -1 \right ] + 1 \right ) d\gamma 
\] \[
  = \int_0^1  \prod_{j=1}^{n-1} \left ( \gamma \left [ X_j \frac{\tilde{j}}{t - \tilde{S}_{j-1}} -1 \right ] + 1 \right ) d\gamma = Y_{n-1}.
\] Hence, under the null hypothesis, \((Y_j )_{j=1}^N\) is a
nonnegative closed martingale with expected value 1, and Kolmogorov's
inequality implies that for any \(J \in \{1, \ldots, N\}\), \[
   \Pr \left ( \max_{1 \le j \le J} Y_j(t) > 1/p \right ) \le p.
\] This method for finding a \(P\)-value for the hypothesis
\(\bar{A}^b \le 1/2\) is also implemented in the SHANGRLA software,
using a novel approach to integrating the polynomial recursively due to
Steven N. Evans (U.C. Berkeley). 
This was the risk-measuring function
used in the San Francisco IRV process pilot audit in November, 2019.

\hypertarget{sec:ballot-comparison}{%
\subsection{Ballot-comparison audits}\label{sec:ballot-comparison}}

Ballot-comparison audits require the voting system to export a
\emph{cast vote record} (CVR) for each physical ballot---the system's interpretation 
of that ballot---in such a way that the corresponding physical ballot can be retrieved,
and vice versa.\footnote{%
However, see section sec.~\ref{sec:zombie}
below.
} 
Suppose that we apply the assorters for a contest to the CVRs
(rather than the actual physical ballots),
and every assorter has an average greater than 1/2, i.e.,
the assertions are true for the CVRs.
Then the assertions are true for
the physical ballots provided the CVRs did not inflate the average value
of the assorter by more than the \emph{assorter margin}, twice the mean
of the assorter applied to the reported CVRs, minus 1, as we shall see.

By how much could error in an individual CVR inflate the value of the
assorter compared to the value the assorter has for the actual ballot?
Since the assorter does not assign a negative value to any ballot, the
\emph{overstatement error} \(\omega_i\) for CVR \(i\) is at most the
value the assorter assigned to CVR \(i\).

The existence of an upper bound for the overstatement error is key to
auditing: otherwise, a single extreme value could make a contest result
wrong, and it would take a prohibitively large sample to rule out that
possibility.

If we can reject the hypothesis that the mean overstatement error
for an assorter is large
enough to account for the assorter margin,
we may conclude that the assertion that the assorter mean exceeds 1/2 is true,
and the audit of that assertion can stop.

Let \(b_i\) denote the \(i\)th ballot, and let \(c_i\) denote the
cast-vote record for the \(i\)th ballot. Let \(A\) denote an assorter,
which maps votes on a ballot card or on a CVR into \([0, u]\), where
\(u\) is an upper bound on the value \(A\) assigns to any ballot card or
CVR. For instance, for plurality voting, \(u=1\); for super-majority,
\(u = 1/(2f)\), where \(f\) is the fraction of valid votes required to
win.

The \emph{overstatement error} for the \(i\)th ballot is
\begin{equation} 
   \omega_i \equiv A(c_i) - A(b_i) \le A(c_i) \le u.
\label{eq:overstatement}\end{equation}
It is the amount by which the assorter applied to the
cast vote record overstates the value of the assorter applied to
corresponding physical ballot.

Let \[
    \bar{A}^c \equiv \frac{1}{N} \sum_{i=1}^N A(c_j)
    \mbox{ and } 
    \bar{\omega} \equiv \frac{1}{N} \sum_{i=1}^N \omega_j.
\] Now \(\bar{A}^b = \bar{A}^c - \bar{\omega}\), so \(\bar{A}^b > 1/2\)
iff \(\bar{\omega} < \bar{A}^c - 1/2\). We know that \(\bar{A}^c > 1/2\)
(or the assertion would not be true for the CVRs), so
\(\bar{\omega} < \bar{A}^c - 1/2\) iff \[
  \frac{\bar{\omega}}{2\bar{A}^c - 1} < 1/2.
\] Define \(v \equiv 2\bar{A}^c - 1\), the \emph{reported assorter
margin}. In a two-candidate plurality contest, \(v\) is the fraction of
ballot cards with valid votes for the reported winner, minus the
fraction with valid votes for the reported loser. This is the
\emph{diluted margin} of \cite{stark10d,lindemanStark12}. 
(Margins
are traditionally calculated as the difference in votes divided by the number
of valid votes. \emph{Diluted} refers to the fact that the denominator
is the number of ballot cards, which is greater than or equal to the
number of valid votes.)

With this notation, the condition for the assertion to be true is: \[
 \frac{\bar{\omega}}{v} < 1/2.
\] Let \(\tau_i \equiv 1- \omega_i/u \ge 0\), and
\(\bar{\tau} \equiv (1/N) \sum_{i=1}^N \tau_i = 1 - \bar{\omega}/u\).
Then \(\bar{\omega} = u(1-\bar{\tau})\) and \[
\frac{\bar{\omega}}{v} = \frac{u}{v} (1-\bar{\tau}).
\] Now \(\omega/v < 1/2\) iff \(\frac{u}{v}(1- \bar{\tau}) < 1/2\),
i.e., \[
    -\frac{u}{v}\bar{\tau}  < 1/2-\frac{u}{v}
\] \[
   \bar{\tau} > 1 - \frac{v}{2u}
\] \[
   \frac{\bar{\tau}}{2 - \frac{v}{u}} > 1/2.
\] 
Finally, define
\[
B(b_i, c) \equiv \tau_i/(2-v/u) = \frac{1-\omega_i/u}{2-v/u} > 0, \;\;
i=1, \ldots, N.
\]
 Then \(B\) assigns nonnegative numbers to ballots,
and the outcome is correct iff \[
   \bar{B} \equiv \frac{1}{N} \sum_{i=1}^N B_i > 1/2.
\] It is an assorter! Any technique that can be used with ballot
polling, including those in sec.~\ref{sec:ballot-polling}, can also be
used to test the assertion \(\bar{B} > 1/2\).

This assertion-based approach to comparison audits is sharper than
methods that rely on the maximum across-contest relative overstatement
of margins (MACRO) \cite{stark10d} in at least two ways: avoiding combining
overstatements across candidates or contests gives a sharper upper bound
on the total error in each (winner, loser) sub-contest, and the test
statistic avoids combining overstatements across contests, which
otherwise can lead to unnecessary escalation of the audit if an
overstatement is observed in a contest with a wide margin.

\hypertarget{sec:stratified}{%
\subsection{Stratified audits}\label{sec:stratified}}

Stratified sampling can be useful within a jurisdiction
if the jurisdiction has a heterogenous
mix of election equipment with different capability, e.g.,
ballot-polling for precinct-based optical scan, where it can be
difficult to associate ballot cards with cast-vote records, and
ballot-level comparisons for central-count optical scan.
Stratification can also be useful auditing contests
that cross jurisdictional boundaries by allowing those
jurisdictions to sample independently from each other.

Stratified sampling for RLAs has been addressed in a number of papers,
including \cite{stark08a,stark08d,stark09a,higginsEtal11,ottoboniEtal18}. 
The central idea of the approach taken in
SUITE \cite{ottoboniEtal18} can be used with SHANGRLA to accommodate
stratified sampling and to combine ballot-polling and ballot-level
comparison audits: Look at all allocations of error across strata that
would result in an incorrect outcome. Reject the hypothesis that the
outcome is incorrect if the maximum \(P\)-value across all such
allocations is less than the risk limit.

SHANGRLA will generally yield a sharper (i.e., more efficient) test than
SUITE, because it deals more efficiently with ballot cards that do not
contain the contest in question, because it avoids combining
overstatements across candidate pairs and across contests, and because
it accommodates sampling without replacement more efficiently.

With SHANGRLA, whatever the sampling scheme used to select ballots or
groups of ballots, the underlying statistical question is the same: is
the average value of each assorter applied to all the ballot cards
greater than 1/2?

Suppose the cast ballot cards are partitioned into \(S \ge 2\)
\emph{strata}, where stratum \(s\) contains \(N_s\) ballot cards, so
\(N = \sum_{s=1}^S N_s\). Let \(\bar{A}_s^b\) denote the mean of the
assorter applied to just the ballot cards in stratum \(s\). 
Then \[
   \bar{A}^b = \frac{1}{N} \sum_{s=1}^S N_s \bar{A}_s^b = \sum_{s=1}^S \bar{A}_s^b \frac{N_s}{N}.
\] 
We can reject the hypothesis \(\bar{A}^b \le 1/2\) if we can reject
the hypothesis 
\[
   \bigcap_{s \in S} \left \{ \bar{A}_s^b \frac{N_s}{N}  \le \beta_s \right \}
\] for all \((\beta_s)_{s=1}^S\) such that
\(\sum_{s=1}^S \beta_s \le 1/2\). Let \(P_s(\beta_s)\) be a \(P\)-value
for the hypothesis \(\bar{A}_s^b \le \frac{N}{N_s}\beta_s\). That
\(P\)-value could result from ballot polling, ballot-level comparison,
batch-level comparison, or any other valid method. For instance, it
could be produced by the methods described in
sec.~\ref{sec:ballot-polling} or sec.~\ref{sec:ballot-comparison}.

Suppose that the samples from different strata are independent, so that
\(\{P_s(\beta_s) \}_{s=1}^S\) are independent random variables. Then
Fisher's combining function (or any other method for nonparametric
combination of tests) can be used to construct an overall \(P\)-value
for the intersection null hypothesis
\(\bigcap_{s \in S} \left \{ \frac{N_s}{N} \bar{A}_s^b \le \beta_s \right \}\).
In particular, if the intersection hypothesis is true, then the
probability distribution of \[
   -2 \sum_{s=1}^S \ln P_s(\beta_s)
\] is dominated by the chi-square distribution with \(2S\) degrees of
freedom, as discussed in \cite{ottoboniEtal18}. 
That makes it possible
to assign a conservative \(P\)-value to the hypothesis
\(\bigcap_{s \in S} \left \{ \frac{N_s}{N} \bar{A}_s^b \le \beta_s \right \}\)
for every \((\beta_s)_{s=1}^S\) such that
\(\sum_{s=1}^S \beta_s \le 1/2\). If all such \(S\)-tuples can be
rejected, we may conclude that \(\bar{A}_b > 1/2\).

\hypertarget{sec:zombie}{%
\subsection{Zombie Bounds II: Return of the Missing
Ballot}\label{sec:zombie}}

\cite{banuelosStark12} discuss how to conduct RLAs when not every
ballot is accounted for or when a ballot cannot be retrieved. They cover
both ballot-polling audits and ballot-level comparison audits. This
section presents a brief but more systematic treatment for ballot-level
comparison audits, reflecting what the SHANGRLA software implements.
This method also makes it possible to use ballot-card style information
to target the sampling to ballot cards that the voting system claims
contain the contest, while protecting against the possibility that the
voting system does not report that information accurately.

To conduct a RLA, it is crucial to have an upper bound on the total
number of ballot cards cast in the contest. Absent such a bound,
arbitrarily many ballots could be missing from the tabulation, and the
true winner(s) could be any candidate(s). Let \(N\) denote an upper
bound on the number of ballot cards that contain the contest. Suppose
that \(n \le N\) CVRs contain the contest and that each 
of those CVRs is
associated with a unique, identifiable physical ballot card that can be
retrieved if that CVR is selected for audit. The phantoms-to-evil
zombies approach is as follows.

If \(N > n\), create \(N-n\) ``phantom ballots'' and \(N-n\) ``phantom
CVRs.'' Calculate the assorter mean for all the CVRs---including the
phantoms---treating the phantom CVRs as if they contain no valid
vote in the contest contest 
(i.e., the assorter assigns the value \(1/2\) to phantom CVRs).
Find the corresponding assorter margin (twice the assorter mean, minus
1).

To conduct the audit, sample integers between 1 and \(N\).

\begin{itemize}
\tightlist
\item
  If the resulting integer is between \(1\) and \(n\), retrieve and
  inspect the ballot card associated with the corresponding CVR.

  \begin{itemize}
  \tightlist
  \item
    If the associated ballot contains the contest, calculate the
    overstatement error as in equation \{eq.~\ref{eq:overstatement}\}.
  \item
    If the associated ballot does not contain the contest, calculate the
    overstatement error using the value the assorter assigned to the CVR, but
    as if the value the assorter assigns to the physical ballot is zero
    (that is, the overstatement error is equal to the value the assorter
    assigned to the CVR).
  \end{itemize}
\item
  If the resulting integer is between \(n+1\) and \(N\), we have drawn a
  phantom CVR and a phantom ballot. Calculate the overstatement error as if
  the value the assorter assigned to the phantom ballot was 0 (turning
  the phantom into an ``evil zombie''), and as if the value the assorter
  assigned to the CVR was 1/2.
\end{itemize}

\textbf{Proposition:} if the risk is calculated based on this
substitution of ``evil zombies'' for ``phantoms,'' the result is still a
RLA with risk limit $\alpha$.

\textbf{Proof:} Every unaccounted for ballot card that might have or
should have contained the contest is treated in the least favorable way.
Every unaccounted for CVR is treated in exactly the way it was tabulated
by the assorter, namely, it is assigned the value 1/2.

Some jurisdictions, notably Colorado, redact CVRs if revealing them
might compromise vote anonymity. If such CVRs are omitted from the tally
and the number of phantom CVRs and ballots are increased
correspondingly, this approach still leads to a valid RLA. But if they
are included in the tally, then if they are selected for audit
they should be treated as if they had the
value \(u\) (the largest value the assorter can assign) in calculating
the overstatement error.

\hypertarget{sec:discussion}{%
\section{Discussion}\label{sec:discussion}}

\hypertarget{sec:multiplicity}{%
\subsection{From many, one}\label{sec:multiplicity}}

Even though SHANGRLA may involve testing many assertions in the audit of
one or more contests, there is no need to adjust for
\emph{multiplicity}. If any assertion is false, the chance that its
complementary hypothesis will be rejected is at most \(\alpha\). If more
than one assertion is false, the chance that all the complementary
hypotheses will be rejected is at most \(\alpha\), because the
probability of the intersection of a collection of events cannot exceed
the probability of any event in the collection. Thus, if any of the
reported winners did not really win, the chance that every complementary
null hypothesis will be rejected is at most \(\alpha\): the chance that
the audit will stop without a full hand count is not greater than the
risk limit.

\hypertarget{sharpness-and-efficiency}{%
\subsection{Sharpness and Efficiency}\label{sharpness-and-efficiency}}

Extant comparison audit methods rely on MACRO, the maximum
across-contest relative overstatement of margins \cite{stark08d}. 
MACRO
is embedded in Colorado's CORLA audit tool, in the Arlo audit tool, and
\url{https://www.stat.berkeley.edu/~stark/Vote/auditTools.htm}.
MACRO involves combining discrepancies across pairs of candidates and
across contests in a way that is conservative, but not sharp. That is,
the condition that is tested is \emph{necessary} for one or more reported
outcomes to be incorrect, but not \emph{sufficient}. 
In contrast, by keeping
the pairwise margins separate, SHANGRLA is sharp for plurality,
super-majority, approval, Borda, STAR-Voting, D'Hondt, etc.---but in general not
for RCV/IRV.
The conditions it tests are both necessary and sufficient
for one or more outcomes to be incorrect. 
This generally allows smaller
sample sizes to confirm the results when the reported contest outcomes
are correct.

\hypertarget{the-power-of-positivity}{%
\subsection{The Power of Positivity}\label{the-power-of-positivity}}

Working with assertions reduces election auditing to testing hypotheses
of the form \(\bar{A}^b < 1/2\): the only statistical issue is to test
whether the mean of a finite list of nonnegative numbers is less than
1/2. As new techniques for testing that hypothesis are developed, they
can be applied immediately to election audits.

\hypertarget{to-halve-or-halve-not}{%
\subsection{To Halve or Halve not?}\label{to-halve-or-halve-not}}

Assertions might look more elegant expressed as \(\bar{A}^b > 1\) rather
than \(\bar{A}^b > 1/2\), which would just involve re-defining \(A\) by
multiplying it by 2. 
However, keeping the connection between assorter
means and getting more than 50\% of the vote in a two-candidate majority
contest seems to be a helpful mnemonic. It also might feel more natural
to write an assertion as or \(\bar{A}^b > 0\), but that would cut the
connection to getting more than a 50\% vote share and make the
lower bound less natural than the nonnegativity constraint
\(A(b_i) \ge 0\) for all \(i\).

Similarly, defining the ``assorter margin'' to be
\(v \equiv 2(\bar{A}^c - 1/2)\) rather than \(\bar{A}^c - 1/2\) keeps
the parallel to a two-candidate plurality contest, where the ``margin''
would generally be defined to be the winner's vote share minus the
loser's vote share.

\hypertarget{sec:conclusions}{%
\section{Conclusions}\label{sec:conclusions}}

Risk-limiting audits of a broad variety of social choice functions can
be reduced to testing whether the mean of any list among a set of finite
lists of nonnegative numbers is less than or equal to 1/2. That is,
Sets of Half-Average Nulls Generate Risk-Limiting Audits (SHANGRLA).
Those hypotheses can be tested directly, e.g., by ballot polling, or
indirectly, by ballot-level comparisons or other methods. They can also
be tested using Bernoulli sampling \cite{ottoboniEtal19}, stratified
sampling, and ``hybrid'' methods following the same general approach as
SUITE \cite{ottoboniEtal18} (see \ref{sec:stratified}), but SHANGRLA is
generally more efficient. The sampling unit can be an individual ballot,
or a cluster of ballots (e.g., all ballots cast in a single precinct or
tabulated by a particular machine).
Samples can be drawn with or without replacement.

Ballot-level comparison audits can also be framed as testing whether any
of the means of a set of finite lists of nonnegative numbers is less
than or equal to 1/2, allowing exactly the same statistical tests to be
used for ballot-polling audits and for ballot-level comparison audits,
unifying the treatment of audits.

This paper proves the validity of two hypothesis tests for that
statistical problem based on sampling without replacement, both of which
were stated without proof in a now-defunct website of Harold Kaplan but
apparently never published. Both proofs are based on Kolmogorov's
inequality for optionally stopped martingales.

Even though auditing one or more contests generally involves testing
many half-average nulls, no multiplicity adjustment is needed, because
the audit only stops if all the nulls are rejected.

For many social choice functions (including plurality, multi-winner
plurality, majority, super-majority, approval, D'Hondt, Borda count, and
STAR-Voting), SHANGRLA comparison audits are sharper than previous
comparison audit methods based on MACRO because the conditions it tests
are both necessary and sufficient for the reported outcomes to be wrong,
while previous methods tested conditions that were necessary but not
sufficient. (MACRO bounds the maximum of a set of sums by the sum of the
term-by-term maxima, both in the condition and in the test statistic;
SHANGRLA keeps the maxima separate, both in the condition and in the
test statistic.)

SHANGRLA also ``plays nice'' with the phantoms-to-zombies approach
\cite{banuelosStark12} for dealing with missing ballot cards and
missing cast-vote records, which has two benefits: (i) it makes it easy
to treat missing ballots rigorously, and (ii) it can substantially
improve the efficiency of auditing contests that do not appear on every
ballot card, by allowing the sample to be drawn just from cards that the
voting system claims contain the contest, without having to trust that
the voting system correctly identified which cards contain the contest.

Open-source software implementing SHANGRLA comparison audits and the
phantoms-to-zombies approach is available; the software was tested in a
process pilot audit of an IRV contest in San Francisco, CA, in November 2019.

\bibliography{bib.bib}

\end{document}